\documentclass[12pt,preprint]{aastex}

\usepackage{graphicx}
\usepackage{amsmath}
\usepackage{amsfonts}
\usepackage{amssymb}
\usepackage{url}
\usepackage{aas_macros}

\title{Simulations of  Stellar Magnetoconvection using the Radiative MHD Code `StellarBox'}
\author{ Alan A. Wray$^{*}$, Khalil Bensassi$^{*,\dag}$, Irina N. Kitiashvili$^{*}$, Nagi N. Mansour$^{*}$, Alexander G. Kosovichev$^{\ddagger}$}

\shortauthors{Wray et al.}

\shorttitle{Simulations of  Stellar Magnetoconvection}

\affil{$^{*}$NASA Ames Research Center,
 Moffett Field, CA 94035-1000, USA\\
$^{\dag}$University of California, Los Angeles (UCLA ),
Department of Earth, Planetary, and Space Sciences,
595 Charles Young Drive East, Los Angeles, CA 90095-1567, USA\\
$^{\ddag}$
 New Jersey Institute of Technology, Newark, NJ 07102, USA\\
e-mail: Alan.A.Wray@nasa.gov, kbensass@nasa.gov\\
web page: https://www.nas.nasa.gov/hms/}


\begin{abstract}
Realistic numerical simulations, i.e., those that make minimal use of ad~hoc modeling, are essential for understanding the complex turbulent dynamics of the interiors and atmospheres of the Sun and other stars and the basic mechanisms of their magnetic activity and variability. The goal of this paper is to present a detailed description and test results of a compressible radiative MHD code, `StellarBox', specifically developed for simulating the convection zones, surface, and atmospheres of the Sun and moderate-mass stars. The code solves the three-dimensional, fully coupled compressible MHD equations using a fourth-order Pad{\'e} spatial differentiation scheme and a fourth-order Runge-Kutta scheme for time integration. The radiative transfer equation is solved using the Feautrier method for bi-directional ray tracing and an opacity-binning technique. A specific feature of the code is the implementation of subgrid-scale MHD turbulence models. The data structures are automatically configured, depending on the computational grid and the number of available processors, to achieve good load balancing. We present test results and illustrate the code's capabilities for simulating the granular convection on the Sun and a set of main-sequence stars. The results reveal substantial changes in the near-surface turbulent convection in these stars, which in turn affect properties of the surface magnetic fields.  For example, in the solar case initially uniform vertical magnetic fields tend to self-organize into compact (pore-like) magnetic structures, while in more massive stars such structures are not formed and the magnetic field is distributed more-or-less uniformly in the intergranular lanes. 
\end{abstract}

\begin{document}

\section{Introduction} 

During the last few decades, there has been a major increase in the use of high-performance computing in computational physics in general and in space science in particular. Massively parallel algorithms have been developed to implement accurate physical models in order to provide efficient and realistic simulations of astrophysical phenomena, for instance, stellar and solar interior dynamics, which is the focus here. Such computational tools are used to support and analyze ground and spacecraft observations. The coupling between observations and numerical simulations is necessary for improving our understanding of complex phenomena on the Sun and other stars because the two approaches are complementary. Recent high-resolution observations of the Sun have uncovered a rich small-scale dynamics which plays a key role in the observed large-scale phenomena. For example, observations and simulations led to the discovery of small-scale vortex tubes generated in intergranular lanes \citep{Brandt1988,Bonet2008,Bonet2010,Steiner2010}, which are a source of the Sun's acoustic emission and magnetic flux concentrations \citep{Kitiashvili2010,Kitiashvili2011,Kitiashvili2012}. Stellar observations from the Kepler mission and ground-based telescopes have revealed that solar-type acoustic oscillations are common in other stars, giving rise to the rapid development of asteroseismology. Also, these observations have found a broad range of magnetic activity, including magnetic activity cycles, starspots, and surprisingly strong energy release events (`superflares'). Because the structure and dynamics of the surface of stars other than the Sun are not resolved in observations, it is important to develop numerical simulations capable of reproducing realistic stellar conditions. For solar conditions such simulations can be verified by comparing with high-resolution observations, providing confidence for using these simulations in the interpretation of stellar observations.

One particular class of these numerical investigations is the realistic simulation of the upper part of the convection zone and the lower part of the atmosphere of the Sun  \citep{Nordlund1982,Keller2004,Carlsson2004,Vogler2005,Stein2006,Georgobiani2012} and other stars \citep{Giorgobiani2003,Steffen2005,Kupka2009,Steffen2009,Muthsam2011,Georgobiani2012,Ludwig2012,Grimm-Strele2015,Mundprecht2015}.  Such investigations typically consist of compressible radiative MHD simulations in which the governing conservation equations of mass, momentum, energy, and magnetic flux are integrated in time.  The stellar composition is modeled using a specific  tabular equation of state (EOS) and optical opacity, both based on given stellar chemical abundances. The computational cost of this approach is high, and currently only a relatively small part of a stellar body can be simulated with a resolution sufficient to study the turbulent  dynamics of the surface and atmosphere in detail, e.g., to resolve in detail granulation and acoustic events. Nevertheless, these small-scale simulations provide knowledge about the structure and dynamics of the near-surface turbulent layer---one of the most complex regions inside stars. This layer plays a key role in energy and mass transport from a star's interior to its corona and wind, as result of strong coupling among plasma dynamics, radiation, and magnetic fields.  This leads to turbulent convection forming high-speed downdrafts driven by radiative cooling and to the magnetic field becoming organized into strong-field structures that dominate the outer envelope. Knowledge of this dynamics, gleaned from high-resolution local simulations, can then be used in global-star models with relatively low resolution. 

The current paper describes the implementation and testing of the radiative MHD code `StellarBox'. The code provides realistic simulation of solar and stellar convection zones and atmospheres and has been used  for a wide range of problems, such as multi-scale dynamics and self-organization processes in turbulent magneto-convection, acoustic wave excitation, formation of stable magnetic structures (such as pores and sunspots), eruptions, generation of magnetic fields by local dynamos, simulation of specific local conditions (e.g., sunspot umbrae and penumbrae), and interactions of the turbulent surface and subsurface with atmospheric layers \citep{Jacoutot2008,Kitiashvili2009,Kitiashvili2010,Kitiashvili2013} . In addition, the code makes it possible to use as initial conditions various models of interior structure pre-calculated for stars with specific chemical compositions, masses, and rotation rates. For F- and A-type stars it is feasible to extend the computational domain into the radiative zone to study the dynamics of convective overshoot layers between the radiative and convection zones. In this paper we first focus on a description of the code and tests of its numerical methods.  Then, as an example application, we show the code's capabilities in simulating turbulent magnetoconvection on the Sun and several main-sequence, moderate-mass stars. A detailed analysis of the stellar simulations will be presented in a separate paper.

The paper is structured as follows. After a brief description of the code modules and components, the governing equations and the physical models are given. Details of the numerical methods and boundary conditions are described. After this, validation and  scalability tests are provided. Finally, example results from simulations of solar and stellar magnetoconvection are presented.

\section{Code Description} 
StellarBox was developed at the NASA Ames Research Center to solve the three-dimensional, fully coupled compressible radiative  MHD equations  using a fourth-order compact implicit Pad\'e  scheme for spatial differentiation and  a fourth-order Runge-Kutta scheme for time integration. The choice of the numerical scheme is motivated by previous experience in numerical turbulence modeling \citep[e.g.,][]{Miyauchi1993,Chow2003}. The computational domain is a rectangular volume typically encompassing a horizontal span of tens to hundreds of megameters and a vertical span of a few to several tens of megameters. The domain is  discretized using a Cartesian grid. Spatial resolution is typically in the range of 100 km down to 6 km in the horizontal directions.  Arbitrary (user-specifiable) mesh stretching is used in the vertical direction. The vertical grid spacing typically increases with depth and ranges from from 10 km to 100 km. Periodic boundary conditions  are used in the horizontal directions and characteristic boundary conditions are applied in the vertical.   Stellar rotation is accommodated by an f-plane approximation at a user-specified latitude and rotation rate.

The  code  incorporates the following user-selectable  subgrid-scale turbulence models for the transport of heat and momentum: compressible \citet{Smagorinsky1963} model, dynamic Smagorinsky \citep{Moin1991, Germano1991}, and implicit hyperviscosity \citep{Caughey2002}.

The radiative transport equation (RTE) is solved at every full time-step  by using a ray-tracing, long characteristic algorithm along 18 rays. A second-order \citet{Feautrier1964} method is used to solve the RTE along each ray. Frequencies are opacity-binned into logarithmic bins, typically four in number.  At depths at which the medium is optically thick, diffusive radiative transport may be optionally used instead of ray-tracing for improved efficiency and accuracy.  A tabular Equation of State (EOS) \citep{Rogers1996} and opacity are used. 

StellarBox is a massively parallel code that uses algorithms optimized for parallelization. The parallel data structures are automatically configured, for given mesh dimensions and number of processors, to optimize load balancing.  Differentiation and the radiative transfer solution are accomplished by transposing these data structures so that $x$, $y$, or $z$ is memory-resident as needed.

\begin{figure}[h!]
\centerline{\includegraphics[width=\textwidth]{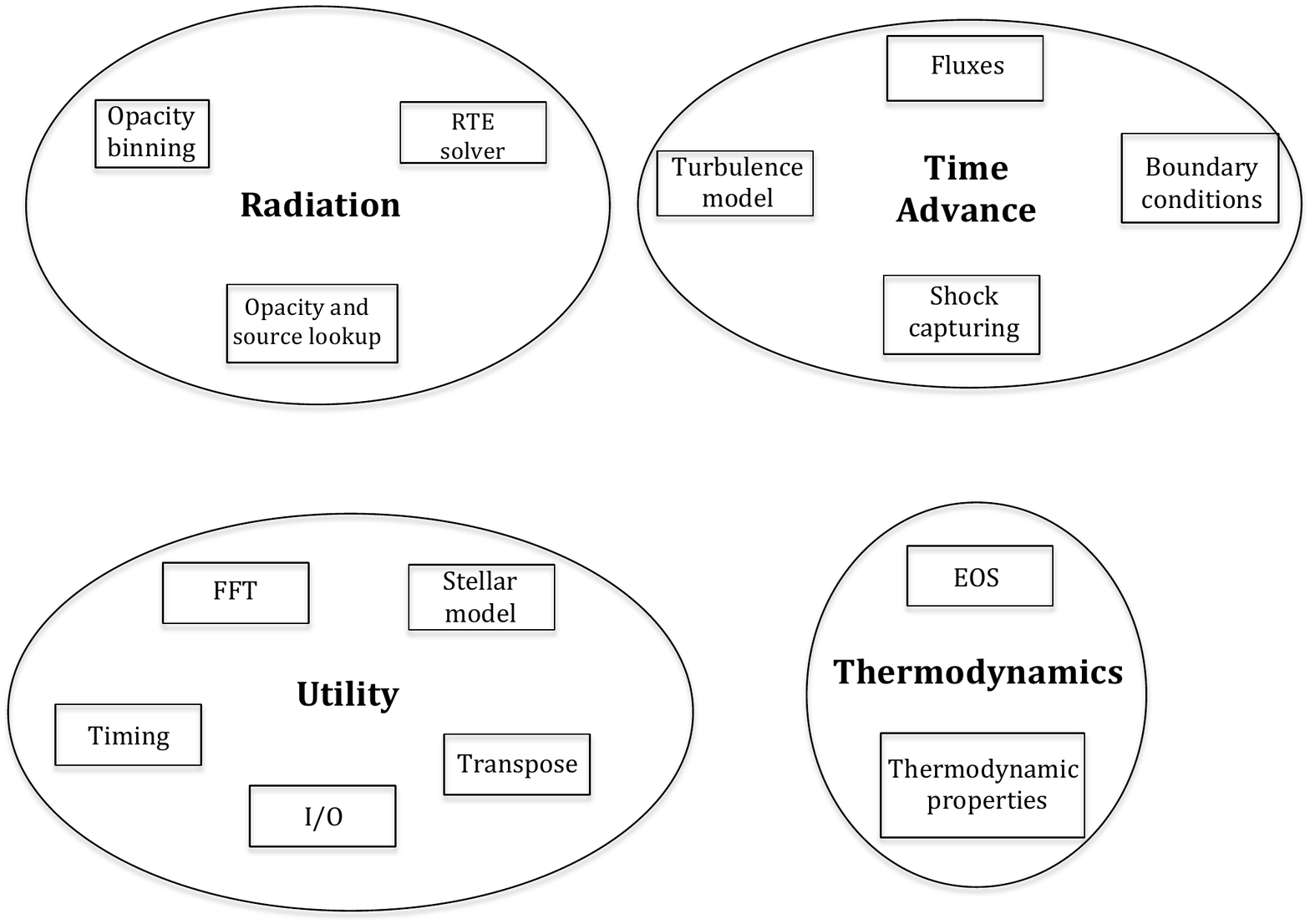}}
\caption{Illustration of the block scheme of the StellarBox code.}
\label{fig:global}
\end{figure}

The overall code structure consists of four main components:  Radiation, Time Advance, Utility, and Thermodynamics. Figure \ref{fig:global} shows the code components and subcomponents.  Their functions are described in the following bullets:
\begin{itemize}
\item \textbf{Radiation}:  implements the opacity-binning model, interpolation algorithms for the opacity and radiation source functions, and the RTE solver.
\item \textbf{Time Advance}: implements the convective and diffusive flux calculations, the shock-capturing and turbulence models, and the boundary conditions to perform the time advance.
\item \textbf{Utility}: implements various utility functions needed by the other modules.
\item \textbf{Thermodynamics}:  implements the equation of state (EOS) and calculates thermodynamic properties.
\end{itemize}

\subsection{Governing Equations}
The equations governing compressible radiative MHD flows are conservation of mass, momentum, energy, and magnetic flux. The code solves grid-cell averaged conservation equations for these quantities.

The  mass conservation equation:
\begin{eqnarray}
\partial_{t}{\rho} + \left(\rho u_{j}\right)_{,j} =0  
\label{eq:MassCons}
\end{eqnarray}
where $\partial_{t}$ denotes the time derivative operator. The subscript $,j$ denotes the space derivative operator  in the $j^{th}$ direction with j =   $\left\{1,2,3\right\}$. Quantity $\rho$  is the averaged mass density and $u_{j} $ is the Favre-averaged, i.e., density-weighted average, velocity component in the $j^{th}$ direction.

 The  momentum conservation equation with gravitational  and  magnetic fields can be written, in the stellar rotation frame, as:
\begin{eqnarray}
\partial_{t}\left(\rho u_{i}\right) +\left(\rho u_{i} u_{j} + p \delta_{ij}\right)_{,j} =\Pi_{ij,j} +{\mathcal{B}}_{ij,j}  - \rho\phi_{,i}-2\epsilon_{ijk}\Omega_j\rho u_k
\label{eq:MomCons}
\end{eqnarray}
where  $p$  is the pressure, $\delta_{ij}$  the Kronecker symbol, $\phi$ the gravitational potential, $\Pi_{ij}$  the viscous  stress tensor, $\epsilon_{ijk}$ the permutation tensor, $\Omega_j$ the stellar mean-rotation vector, and $\mathcal{B}_{ij} $ the magnetic stress tensor
\begin{eqnarray}
\mathcal{B}_{ij} = \frac{1}{4 \pi} B_{i}B_{j} - \frac{1}{8\pi} B_{k}B_{k} \delta_{ij}
\label{eq:Nij}
\end{eqnarray}

The total energy conservation equation reads
\begin{eqnarray}
\partial_{t} E + \left[ \left(E + p \right)u_{j} \right]_{,j} &=& - \left(\phi_{j} u_{i}\right)_{,j}+ \left( {\Pi_{ij}} u_{i}\right)_{,j} + \left({\mathcal{B}_{ij}} u_{i}\right)_{,j}  \nonumber \\ 
&-& {\mathcal{Q}_{j,j}}  - {\mathcal{Q}_{j,j}^{\rm rad}} 
\label{eq:GEtotalenergy}
\end{eqnarray}
where ${\bf{\Pi}}$ is the viscous tensor, $\bf{\mathcal{Q}}$ is the non-radiative heat flux (diffusive and Joule heating),  $\mathcal{Q}_j^{\rm rad}$ is the radiative heat flux, and  $E$ is the total energy per unit volume given  by:
\begin{eqnarray}
E = \rho e +  \frac{1}{2} \rho u_{j} u_{j} + \rho \phi + \frac{1}{8\pi} {B}_{j} {B}_{j}
\label{eq:totalenergy}
\end{eqnarray}
where $e$ is the internal energy per unit mass.  The radiative heat flux is obtained by solving the radiative transfer equation, and the viscous tensor and the non-radiative heat flux are written in terms of transport coefficients and gradients of resolved variables.

The magnetic flux  conservation equation can  be expressed in Gaussian units as 
\begin{eqnarray}
\partial_{t} B_{i} +\left(u_{j}B_{i}-u_{i}B_{j}\right)_{,j} =\left( \dfrac{c^{2}}{4 \pi \sigma} \left[B_{i,j}-B_{j,i} \right]\right)_{,j}
\label{eq:Bconservation}
\end{eqnarray}
where $B_{j}$  is the magnetic flux density in the $j^{th}$ direction, $c$ the speed of light, and $\sigma$ the electrical conductivity.


\subsection{Subgrid Stress Tensor}
Since it is currently impossible to achieve realistic Reynolds numbers in solar and stellar simulations,  StellarBox utilizes a Large-Eddy Simulation  (LES) approach in which models are used to approximate the effect of subgrid motions on the resolved scales. Modeled subgrid quantities will be denoted by a subscript $T$. The subgrid stress tensor is represented in the form:
   \begin{eqnarray}
\Pi_{ij} =  2\mu_{T}\left(S_{ij}- \frac{1}{3} u_{k ,k}\delta_{ij}\right) - k_{T} \delta_{ij}\label{eq:GTau}
\end{eqnarray}
in which $\mu_{T}$ is the subgrid viscosity  given by the \citet{Smagorinsky1963} model, augmented for compressible flows with a shock capturing term:
\begin{eqnarray}
\mu_{T} = \rho \Delta^{2} \left(C_{S} |S| + C_{D} | u_{k,k}  | \right)
\label{eq:mut}
\end{eqnarray}
where $C_{S}$ is a Smagorinsky coefficient, $C_{D}$ is a shock-capturing coefficient, $\Delta$ is the grid spacing, $|S| =  \sqrt{\left(2 S_{ij}S_{ij}\right)}$, and  $S_{ij}$ is the strain-rate tensor
\begin{eqnarray}
S_{ij} =   \frac{1}{2} \left( u_{i,j} +u_{j,i}  \right);
\label{eq:Sij}
\end{eqnarray}
$k_{T}$ is the subgrid kinetic energy density given by
  \begin{eqnarray}
 k_{T} = \frac{2}{3} C_{C} \rho \Delta^{2} |S|^{2}
 \label{eq:ksgs}
 \end{eqnarray}
  where $C_{C}$ is a second Smagorinsky coefficient.  The code provides  a dynamic option for determining $C_{S}$ and $C_{C}$  based on the associated \citet{Germano1991} identities.  Alternatively, they can be specified by the user, as is $C_{D}$. 
 
 \subsection{Heat Fluxes and Turbulent Resistivity}
 The total non-radiative heat flux is the sum of  two contributions: a term due to temperature gradients, that is, Fourier's law, and a second due to Joule heating:
 \begin{eqnarray}
 {\bf{\mathcal{Q}}}_{j}  = \kappa_{T}T_{,j} +  \left(\frac{c}{4\pi}\right)^{2}  \frac{1}{\sigma_{T}}\left(B_{i,j} B_{i} - B_{i} B_{j,i}\right),
 \label{eq:Qj}
 \end{eqnarray}
 where $\kappa_{T}$ is the subgrid heat conductivity 

\begin{eqnarray}
\kappa_{T} = \frac{c_{p} \; \mu_{T}  }{Pr_{T}}
\label{eq:kappaT}
\end{eqnarray}
in which $Pr_{T}$ is the turbulent Prandtl number and $c_{p}$ is the specific heat at constant pressure per unit mass taken from the Equation Of State (EOS) tables. 
The turbulent Prandtl number $Pr_{T}$ can be specified by the user; it is typically taken to be near unity.

The turbulent electrical conductivity, $\sigma_{T}$, in Gaussian units is given by
\begin{eqnarray}
\sigma_{T} =\dfrac{c^{2}}{4 \pi \eta_{T}},
\label{eq:sigmat}
\end{eqnarray}
where $c$ is the speed of light and $\eta_{T}$ is the  turbulent magnetic diffusivity, modeled following \citet{Balarac2010} and \citet{Theobald1994}:
\begin{eqnarray}
\eta_{T}= C_{B} \Delta^{2} \dfrac{| \epsilon_{ijk} B_{k,j}|}{\sqrt{\rho}},
\label{eq:sigmaT}
\end{eqnarray}
where $C_{B}$ a user-specified constant chosen through comparison to fully resolved MHD turbulent simulations.  A typical value is $0.25$.


\subsection{Radiative Heat Flux} \label{sec:rad}
\begin{figure}[h!]
\centerline{\includegraphics[width=0.5\textwidth]{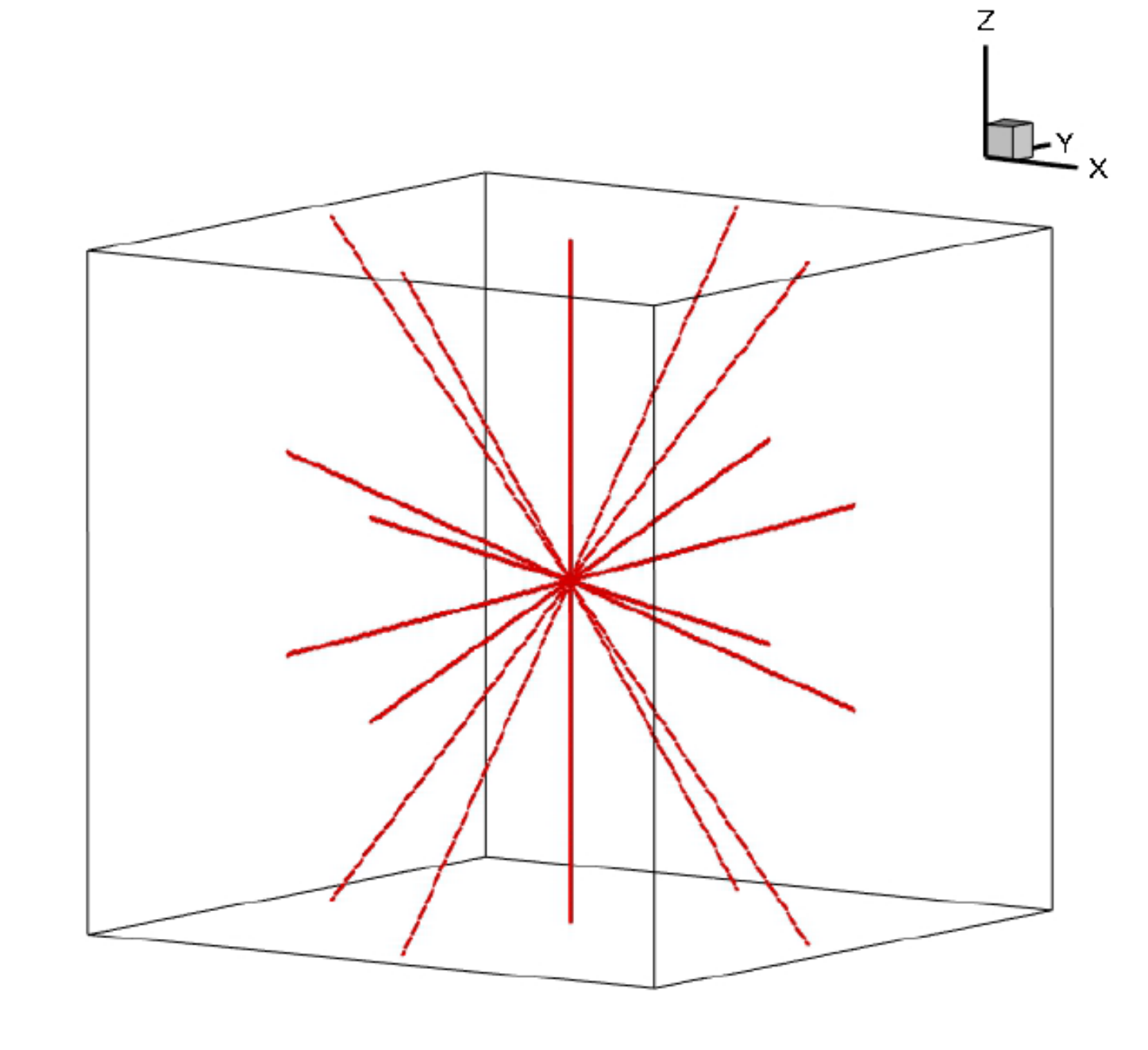}}
\caption{The 18 rays used to cover direction space $\bf{\Omega}$.}
\label{fig:18rays}
\end{figure}

The radiative heat flux is given by

\begin{eqnarray}
{\bf{\mathcal{Q}}}_{j}^{\rm rad} = \int_{0} ^{\infty}\int_{4\pi} \Omega_j \; I_{\nu}\left( \bf{\Omega},\textbf{x} \right)  d^2\Omega d\bf{\nu},
\label{eq:Qrad}
\end{eqnarray}
where the quantity $\bf{\Omega}$ is a direction unit vector, $\nu$ is the frequency, and $I_{\nu}$ is the radiative intensity at frequency $\nu$ given by  the radiative transfer equation:

\begin{equation}
{\bf \Omega} \cdot {\bf \nabla} I_{\nu} \left({\bf \Omega}, {\bf x} , t\right) =\chi_{\nu}\left({\bf x}, t \right)  \left[S_{\nu}\left({\bf x}, t \right) -I_{\nu}\left({\bf \Omega}, {\bf x}, t \right) \right],
\label{eq:RTE}
\end{equation}
where non-isotropic scattering, polarization, and  the finiteness of the speed of light are neglected. $S_{\nu} $ is the radiation source function and   $\chi_{\nu}$ the opacity. Eq. (\ref{eq:RTE}) is solved using the \citet{Feautrier1964} method, which is based on bi-directional ray tracing.  The algorithm implemented in the code uses 18 rays (9 bi-directional rays), shown in Fig. \ref{fig:18rays}.

In the Feautrier  method, the intensity  function along a ray  is divided  into forward- and backward-travelling intensities (${I}^{+}_{\nu}$, ${I}^{-}_{\nu}$ respectively) as a function of distance $s$ along the ray:

\begin{eqnarray}
 \dfrac{dI^{+}_{\nu}}{ds} &=& \chi_{\nu} \left( S_{\nu} -  I^{+}_{\nu} \right)\label{eq:rte1}\\
 \dfrac{dI^{-}_{\nu}}{ds} &= & -\chi_{\nu} \left(  S_{\nu} -  I^{-}_{\nu} \right)
\label{eq:rte2}
\end{eqnarray}
By subtracting Eq. (\ref{eq:rte2}) from Eq. (\ref{eq:rte1}), we obtain 
\begin{eqnarray}
 \dfrac{d \left(I^{+}_{\nu} - I^{-}_{\nu}\right)}{ds} =  2\chi_{\nu}S_{\nu} - \chi_{\nu} \left( I^{+}_{\nu} +I^{+}_{\nu}\right) \label{eq:rte1m}
\end{eqnarray}
and by adding Eq. (\ref{eq:rte1}) and Eq. (\ref{eq:rte2}), we have 
\begin{eqnarray}
 \dfrac{d \left(I^{+}_{\nu} + I^{-}_{\nu}\right)}{ds} =  - \chi_{\nu} \left( I^{+}_{\nu} -I^{+}_{\nu}\right) 
\label{eq:rte2m}
\end{eqnarray}
Defining $I^{D}_{\nu} \equiv I^{+}_{\nu} - I^{-}_{\nu}$ and $ I^{S}_{\nu} \equiv I^{+}_{\nu} + I^{-}_{\nu} 
$ and combining  Eqs.  (\ref{eq:rte1m}) and  (\ref{eq:rte2m}), we obtain a single second-order differential equation

\begin{equation}
\dfrac{d^{2}I^{S}_{\nu}}{ds^{2}} - \frac{1}{\chi_{\nu}}\dfrac{d\chi_{\nu}}{ds} \dfrac{dI^{S}_{\nu}}{ds} - \chi_{\nu}^{2}I^{S}_{\nu}  =  -2  \chi^2_{\nu} S_{\nu}.
\label{eq:rtenew}
\end{equation}
 
 Equation (\ref{eq:rtenew}), as written, is for a monochromatic intensity $I^{S}_{\nu}$.  As mentioned earlier, the code treats frequency space by using an opacity-binning technique, in which the opacity of a given bin $b$, $\chi_b$, is taken to be the Rosseland mean opacity of the frequencies assigned to that bin.  Similarly, the source function $S_b$ for bin $b$ is the source $S_{\nu}$ integrated over those frequencies.  The RTE for the integrated intensity $I_b$ in bin $b$ is then
 
 \begin{equation}
\dfrac{d^{2}I^{S}_b}{ds^{2}} - \frac{1}{\chi_b} \dfrac{d\chi_b}{ds}\dfrac{dI^{S}_b}{ds} - \chi_b^{2}I^{S}_b  =  -2  \chi^2_b S_b
\label{eq:rtebin}
\end{equation}
Eq. (\ref{eq:Qrad}) becomes, expressed in terms of bins
\begin{eqnarray}
{\bf{\mathcal{Q}}}_{j}^{\rm rad} = \sum_{b=1}^{n_b}\int_{4\pi} \Omega_j \; I_b\left( \bf{\Omega},\textbf{x} \right)  d^2\Omega
\label{eq:Qradb}
\end{eqnarray}
where $n_b$ is the number of opacity bins, typically four in running StellarBox.  Upon discretizing $\Omega$, the integral over $\Omega$ in Eq. (\ref{eq:Qradb}) becomes an appropriately weighted sum over the 18 rays shown in Fig. \ref{fig:18rays}.   The weights are chosen to maximize the Taylor-series order of accuracy of the solid-angle integration.
 

\subsection{Numerical Methods}

The computational domain is a rectangular volume of dimensions $L_{x} \times L_{y} \times L_{z}$ defined by the user.  The volume is discretized  into a cartesian grid, where the number of grid points along each axis, $n_{x}$, $n_{y}$, and $n_{z}$,  is set by the user as well.  By convention, negative values of $z$ correspond to locations below the nominal photosphere and positive values above it.

The basic finite difference differentiation scheme is 4th-order Pad\'e and is used to compute derivatives appearing in the diffusive fluxes and also the derivatives of the convective and diffusive fluxes themselves in equations (\ref{eq:MassCons}), (\ref{eq:MomCons}), (\ref{eq:GEtotalenergy}), and (\ref{eq:Bconservation}). The scheme must of course be supplemented with  boundary conditions, as described in the next section.

The basic stencil to compute the first derivative $\mathcal{F}^{'}$ of a quantity $\mathcal{F}$ in the $j^{th}$ direction ($j=\left\{1,2,3\right\}$) is

\begin{equation}
\frac{1}{4}\mathcal{F}^{'}_{k-1} +  \mathcal{F}^{'}_{k}  + \frac{1}{4} \mathcal{F}^{'}_{k+1} = \frac{3}{4h_{k}^j} \left(\mathcal{F}_{k+1}  - \mathcal{F}_{k-1}  \right) ,
\label{eq:pade}
\end{equation}
where the subscripts are grid point indices in the $j^{th}$ direction and $h_{k}^j$ is a measure of the grid spacing at point $k$ in that direction.
Eq. (\ref{eq:pade}) is all that is required in the periodic $x$ and $y$ directions, where

\begin{eqnarray}
h_{k}^1 \equiv \Delta x= \frac{L_{x}}{n_{x}}   & {\rm and} & 
h_{k}^2 \equiv \Delta y = \frac{L_{y}}{n_{y}}.   
\label{eq:spaedxdy}
\end{eqnarray}

In the vertical direction $z$ ($j=3$), the grid spacing is arbitrary, meaning that $h_{k}^3$ may depend on $k$.  Typically, the $z$-grid is stretched below the photosphere where structures are larger. For one-sided derivatives at the boundaries, Eq. (\ref{eq:pade}) is supplemented, e.g.,  at the bottom in $z$, by the second-order boundary form

\begin{equation}
\frac{1}{2}\left(\mathcal{F}^{'}_{1} +  \mathcal{F}^{'}_{2}\right) = \frac{1}{h_{1}^3} \left(\mathcal{F}_{2}  - \mathcal{F}_{1}  \right) 
\label{eq:padebdy}
\end{equation}
and analogously at $n_z$.  Some $z$-derivatives are computed by specifying derivative values at the boundaries, such as where characteristic boundary conditions are used.
$h_{k}^3$ is defined as

\begin{eqnarray}
h_{k}^3 &=&  \frac{1}{2}\left(z_{k+1} - z_{k-1}\right) \;\; \forall \; k=\{2...n_z-1\} \\
h_{1}^3 &=&  z_{2} - z_{1} \\
h_{n_z}^3 &=&  z_{n_z} - z_{n_z-1} 
\label{eq:hs}
\end{eqnarray}   

The source term in Eq. (\ref{eq:GEtotalenergy}) involves the computation of $\nabla \cdot {\bf \mathcal{Q}}^{\rm rad}$ which, as mentioned in Sec. (\ref{sec:rad}), requires  solving Eq. (\ref{eq:rtebin}). The latter is discretized using second order  finite differences and solved for the quantity $I_b^S-2S_b$ so that the limiting case of an optically thick medium is handled accurately.

The system of equations Eqs. (\ref{eq:pade}) and the discretized form of Eq. (\ref{eq:rtebin}) are tridiagonal and are solved using the Thomas algorithm. The code has various tridiagonal solvers implemented for each  type of boundary condition.

\subsection{Boundary Conditions}
In order to close the system of equations (\ref{eq:MassCons}), (\ref{eq:MomCons}), (\ref{eq:GEtotalenergy}), and (\ref{eq:Bconservation}), boundary conditions are required.
Periodic boundary conditions are applied in the horizontal directions ($j={1,2}$).  For the $z$ direction, the user has a choice of both boundaries closed (impenetrable, i.e., $u_3=0$), or both open, or the top open and the bottom closed.  The open boundaries are implemented by a characteristic method \citep[e.g. ][]{Sun1995}.  This top-open, bottom-closed set of boundary conditions is the one typically used in StellarBox.

To simulate the energy flowing from the stellar core, the bottom boundary condition for the total energy is modified by adding an incoming energy flux per unit area equal to the stellar value.  In addition,
to preserve exact conservation of mass, momentum, energy, and magnetic flux in the domain, inward fluxes equal to the sum of those convected and diffused outward through the top and bottom boundaries are introduced at the bottom boundary as an areal average.  This does not apply to the outgoing radiative flux at the top boundary --- the system is allowed to find its own radiative equilibrium wherein the radiative flux emitted through the top boundary statistically balances the incoming energy flux at the bottom boundary.  The state that attains this last condition serves as a sanity check on the general correctness of the simulation.

\subsection{Time Integration}

The discretized system of equations (\ref{eq:MassCons}), (\ref{eq:MomCons}), (\ref{eq:GEtotalenergy}), and (\ref{eq:Bconservation}) can be written compactly as 

\begin{equation}
 \frac{d{\bf{U}}}{dt}   = \mathcal{R}({\bf{U}})
\label{eq:compacRes}
\end{equation}
where $\mathcal{R}({\bf{U}})$ includes all the spatially discretized terms, and $\textbf{U}$ = $\left(\rho,\rho \textbf{u},  E , \textbf{B}\right)^{T}$ is the vector of conserved variables. Eq. (\ref{eq:compacRes}) is solved  using the following $4^{th}$-order Runge-Kutta scheme:

\begin{equation}
{\bf{U}}^{n+1} = {\bf{U}}^{n} + \frac{\Delta t}{6} \left(k_{1}+ 2k_{2}+ 2k_{3}+ k_{4}\right)
\label{eq:RK1}
\end{equation}
$\Delta t$ is the time step, the superscript $n$ stands for the time step, and the quantities $k_{1}$, $k_{2}$, $k_{3}$, and $k_{4}$ are defined as follows

\begin{eqnarray}
k_{1} &=& \mathcal{R}({\bf{U}}^{n})\\
k_{2} &=& \mathcal{R}  \left({\bf{U}}^{n} + k_{1} \frac{\Delta t}{2}\right)\\
k_{3} &=& \mathcal{R}  \left({\bf{U}}^{n} + k_{2} \frac{\Delta t}{2}\right)\\
k_{4} &=& \mathcal{R}  \left({\bf{U}}^{n} + k_{3} \Delta t \right)
\label{eq:RK2}
\end{eqnarray}

\section{Parallel Scaling}

\begin{figure}[h]
\centerline{\includegraphics[width=\textwidth]{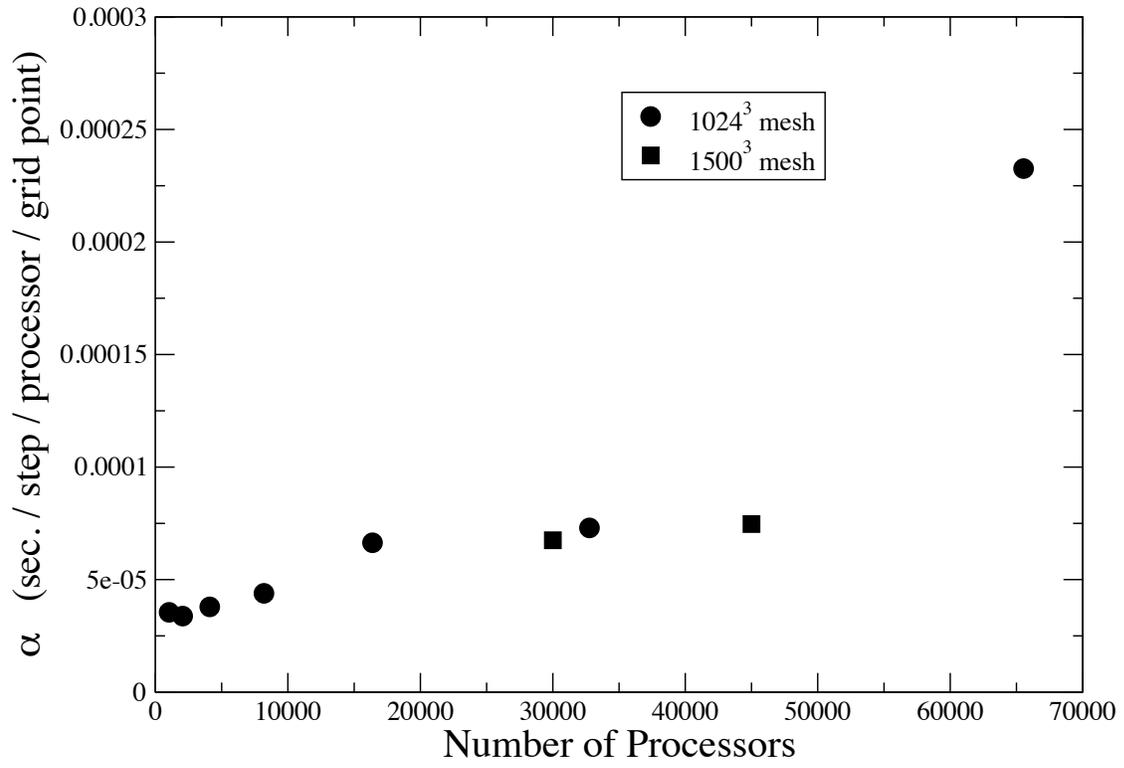}}
\caption{Scaling results for StellarBox: $\alpha$, time per step per processor per grid point, as a function of the number of processors.}
\label{fig:scaling}
\end{figure}

It is essential, for efficient parallel computation, to understand the scaling properties of a massively parallel code such as StellarBox.  Various means of describing these properties have been devised.  Here we present the results
of our scaling tests in a particularly simple format based on comparison to an ideally performing code and computer system.  By ``ideal" we mean that the time associated with communication among the processors increases no faster than the amount of data communicated.  Since all the significant computation in StellarBox involves an equal amount of work for each processor and for each grid point, an expression for the time, $t_{\rm ideal}$, for such an ideal system to compute one time step would have the following form:

\begin{equation}
 t_{\rm ideal}=\alpha\frac{n_x n_y n_z}{N_{\rm proc}}
 \label{scaling_law}
 \end{equation}
 
 \noindent in which $n_x$, $n_y$, $n_z$ are the mesh dimensions and $N_{\rm proc}$ is the number of processors.  The factor $\alpha$, which is the time per step per processor per grid point, is constant in the ideal case for any mesh dimensions or number of processors.  Of course, any real code and computer system will show performance degradation, reflected by an increasing $\alpha$, as  the number of processors increases for a fixed problem size, principally due to contention for limited resources such as inter-processor communication hardware.  In Figure \ref{fig:scaling} we show the $\alpha$ values obtained for various numbers of processors on two different mesh sizes: $1024^3$ and $1500^3$; all runs were conducted on the Pleiades computer system at NASA Ames Research Center.  It is clear from this figure, for meshes of these approximate sizes, that the code behaves well --- only slowly growing $\alpha$ --- up to 45,000 processors or so, and that performance is seriously degraded for 65,000 processors with a mesh of $1024^3$, on this computer system.  Larger mesh dimensions would presumably result in better performance for 65,000 processors and beyond, though we have not tested this at this time.

\section{Test Cases}

SolarBox has been exercised for a large number of test cases to ensure physical and numerical accuracy.  A subset of the more interesting tests is given in the subsections to follow.

\subsection{Sod Shock Tube }
The Sod shock tube problem \citep{Sod1978} is often used as a one-dimensional  test case to check the ability of a compressible code to capture shocks, contact surfaces, and rarefaction waves present in the flow. The initial conditions are:

\[ (\rho,u,p)_{t=0} = \left\{ 
  \begin{array}{l l}
    (0.125,0,0.1) & \quad \text{if $x$ $\le$ 0.5}\\
    (1.,0,1.) & \quad \text{if $x$ $>$ 0.5}
  \end{array} \right.\]

\begin{figure}[h]
\centerline{\includegraphics[width=\textwidth]{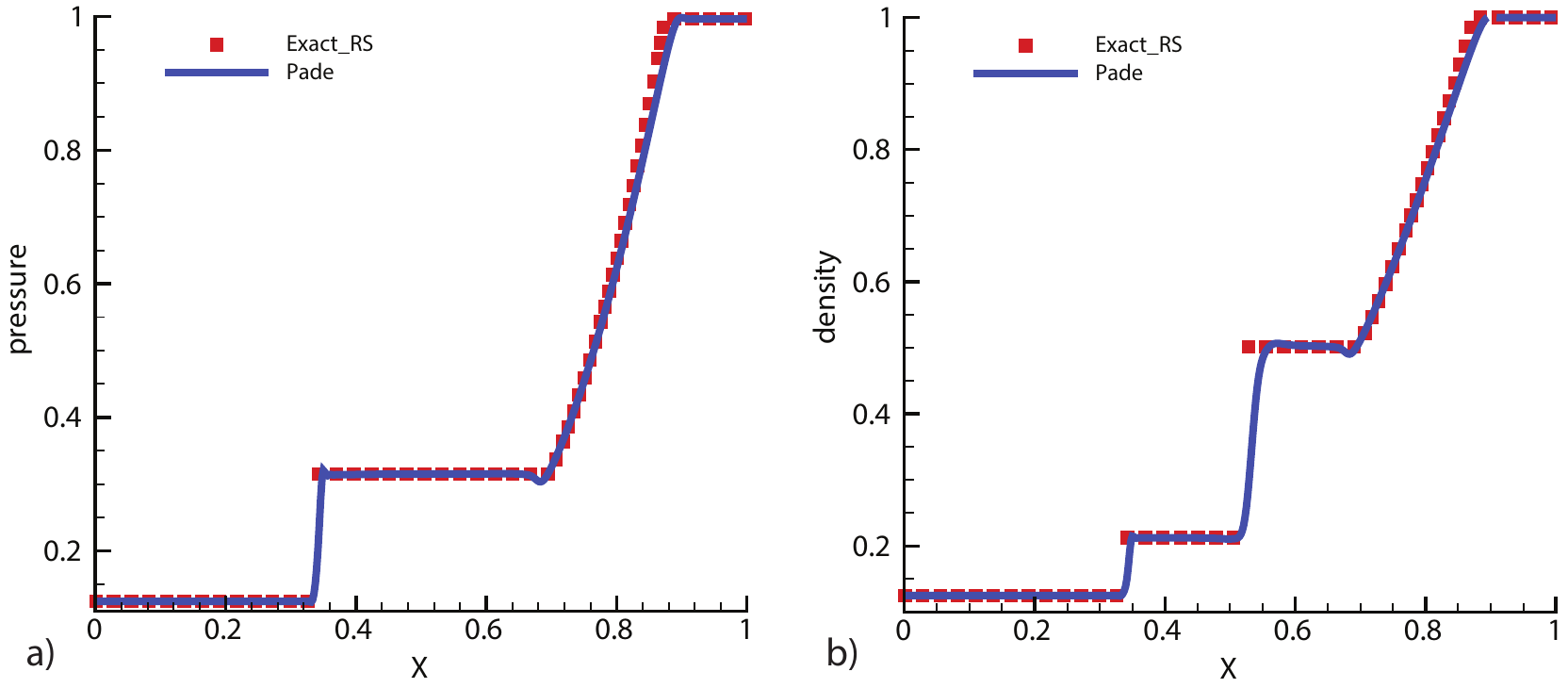}}
\caption{Sod shock-tube test: a) Pressure and b) density profiles at $t= 0.1\,$. Solid line is the numerical solution and symbols represent the exact Riemann solver solution (Exact$_{}$RS).  }
\label{fig:sodP}
\end{figure}

The gas is taken to be perfect with a specific heat ratio $\gamma = 5/3$. Different runs were performed using different grid resolutions and different numerical diffusion coefficients in order to find a balance between damping numerical instabilities and resolving the sharp discontinuities.  In the case of the Sod shock tube, and as a general rule, less numerical diffusion is required for 1D, non-radiative, non-stratified problems than for stellar simulations.  Figure \ref{fig:sodP}  shows  the numerical solution for pressure and density profiles along the $x$ axis at time $t=0.1$, using 480 grid points. The solid line is the numerical solution obtained using the differentiation, shock capturing, and time-advance algorithms in the code, and the symbols show the solution provided by an exact Riemann solver \citep{Toro1997}. Good agreement is obtained for both the positions and the amplitudes of all the flow structures (shock, contact surface, and expansion fan).

\subsection{Orszag-Tang Problem}
The \citet{Orszag1979} problem is a popular  MHD  test for  two dimensional codes and is also known as  the $\nabla\cdot {\bf B}=0$ test condition. It is used to check the robustness of a code in handling MHD shocks, including shock-shock interactions.  The initial conditions  are:

\begin{eqnarray}
\rho = \frac{25}{36 \pi}; & & p=  \frac{5}{12 \pi} \\
u = - \sin(2 \pi y); & & v=  \sin(2 \pi x) \;\;\;\;\;\ \forall x,y \in [0,1] 
\times [0,1]\\
B_{x} = -\sin(2 \pi y); & &  B_{y} = \sin(4 \pi x) \;\;\;\  \forall x,y \in [0,1]\times [0,1]  
\label{eq:OT_{init}}
\end{eqnarray}

\begin{figure}[h]
\centerline{\includegraphics[width=1\textwidth]{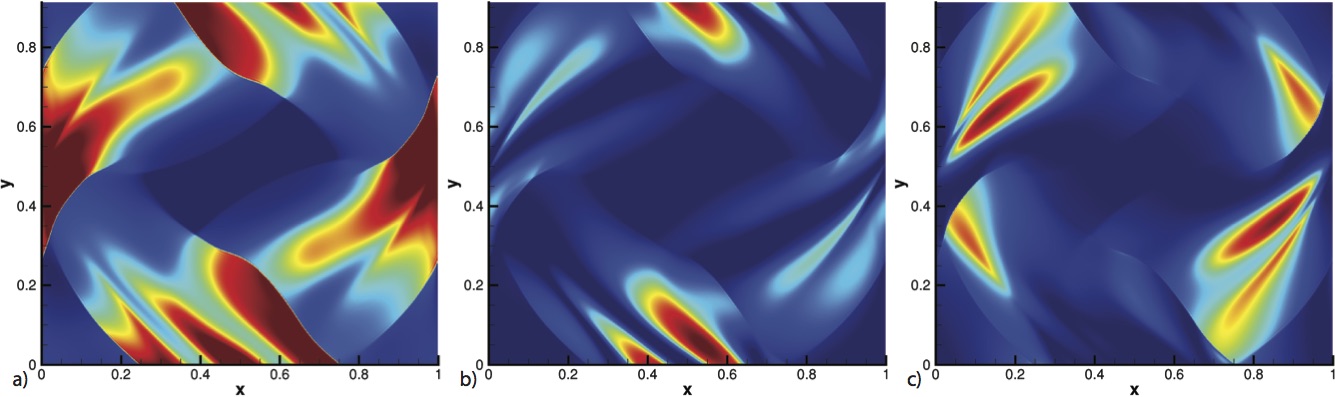}}
 \caption{Solution of the Orszag-Tang problem at $t=0.25$, obtained with the StellarBox using the Pad\'e scheme: a) density (the linear color  map ranges from 0.06 to 0.52), b) magnetic energy  (from 2.$\times10^{-8}$ to 0.3), c) kinetic energy (from 2.$\times10^{-8}$ to 0.65). }
 \label{fig:otrho}
\end{figure}

The computation was performed using a $512\times 512$ grid. At time $t=0.25$,  Figure \ref{fig:otrho} shows snapshots of the density, magnetic energy, and kinetic energy fields. The images  are directly comparable with ones in \citet{Stone2008} and  good quantitative agreement is obtained  even though  the numerical methods used in the two codes are quite different. 

\begin{figure}[h]
\centerline{\includegraphics[width=1\textwidth]{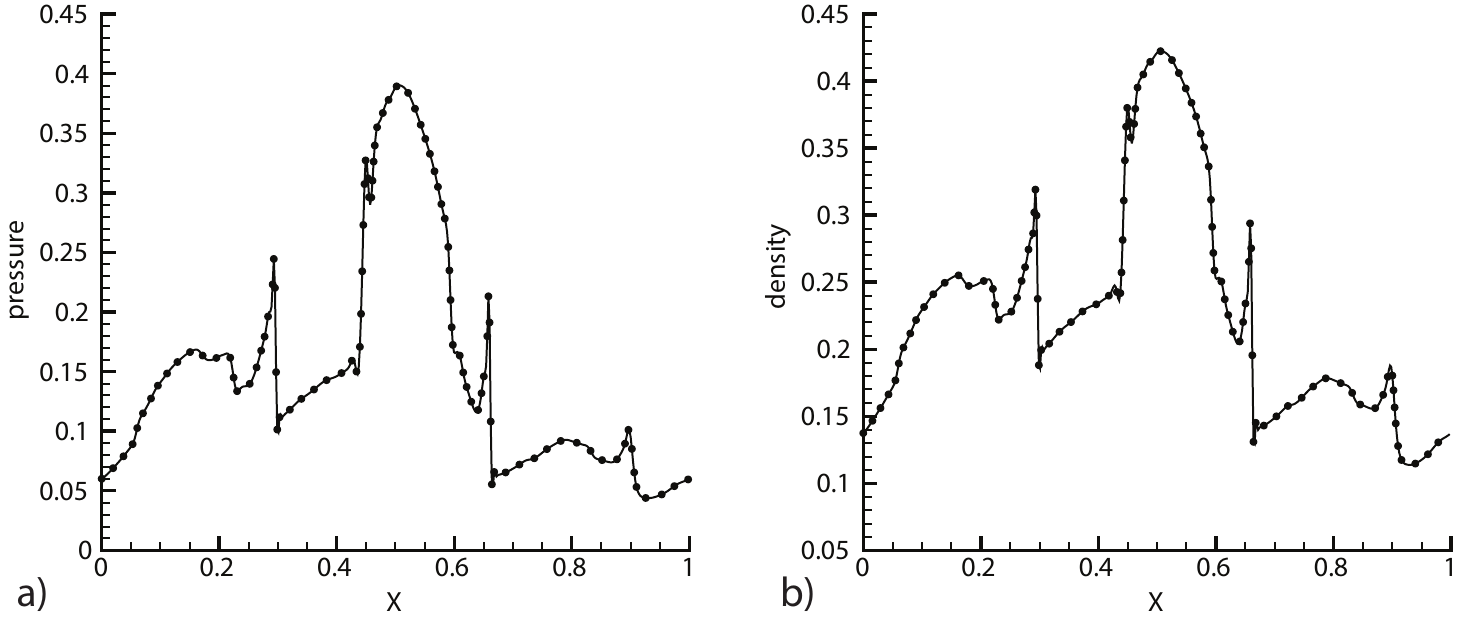}}
 \caption{Solution of the Orszag-Tang problem at $t=0.25$, obtained with StellarBox: a) density and b) pressure profiles $y=0.4277$.}
 \label{fig:otrho1}
\end{figure}

 Figure \ref{fig:otrho1}  shows the pressure and density profiles for $t=0.25$ at $y=0.4277$. These results compare very well with those in \citet{Londrillo2000}, \citet{Jiang1999}, and \citet{Ryu1998}.

\subsection{Brio and Wu Shock Tube }
\begin{figure}[h]
\centerline{\includegraphics[width=\linewidth]{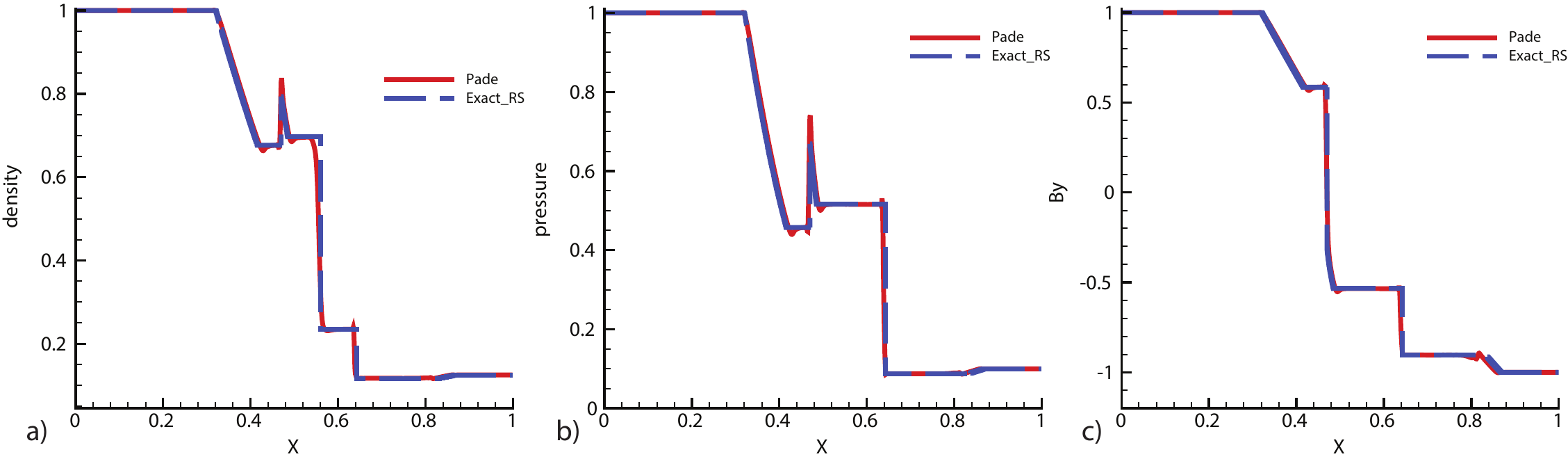}}
  \caption{Brio and Wu shock-tube test at $t=0.1$: the solid line is the numerical result obtained with the Pad\'e scheme, and the dashed line is the solution provided by the exact MHD Riemann solver of \citet{Torrilhon2003}: a) density profile, b)  pressure profile, c) magnetic filed profile.}
    \label{fig:bwfig}
\end{figure}

This test is a simulation of an MHD shocktube \citep{Brio1988}; the hydrodynamic portion of the initial conditions are the same as for the Sod shock tube problem. However, the B field  makes the algebraic equations of  the Riemann problem  highly non-linear and complex in a five-dimensional parameter space. Moreover, the presence of so-called non-regular waves  in the MHD system causes the Riemann problem to be non-unique in some cases \citep{Torrilhon2003}.  The right and left states are initialized as follows: 

\[ (\rho,u,v,B_{y},B_{z},p)_{t=0} = \left\{ 
  \begin{array}{l l}
    (0.125,0,0,1,0,0.1) & \quad \text{if $x$ $\le$ 0.5}\\
    (1,0,0,-1,0,1) & \quad \text{if $x$ $>$ 0.5}
  \end{array} \right.\]
where  $B_{x}$=0.75  is constant and $\gamma$ = 2. 
The numerical results are compared to the solution provided by the exact MHD Riemann solver (Exact$_{-}$RS)  of \citet{Torrilhon2003}.
Figure \ref{fig:bwfig} shows the density, pressure, and $B_{y}$ profiles at $t=0.1$.  Good agreement is obtained for both the regular and non-regular waves.

\subsection{Radiative Transfer Test}
In order to test  the ray tracing algorithm that is used in StellarBox to solve the radiative transfer equation (Eq. \ref{eq:RTE}), two three-dimensional radiative transfer computations were performed using a box of 6$\times$6$\times$6 Mm that includes 1 Mm of the lower solar atmosphere; the grid dimensions were 64$\times$64$\times$128. In the first computation, the ray tracing algorithm was used everywhere; in the second, it was only used in the part of the domain defined by $z \gtrsim -2$~Mm (as previously mentioned,  by convention negative values of $z$ correspond to locations below the nominal photosphere and positive values above it), and a purely diffusive treatment was used everywhere else ($z \lesssim -2$~Mm). The latter is a good approximation in optically thick regions, which is the case below $z \approx-2$~Mm.  The ray-tracing portion was computed using four opacity bins as usual, as described above, and the diffusive portion was computed using a single, all-frequency Rosseland-average opacity value at each point to obtain the diffusion coefficient.

\begin{figure}[h]
\centerline{\includegraphics[width=\linewidth]{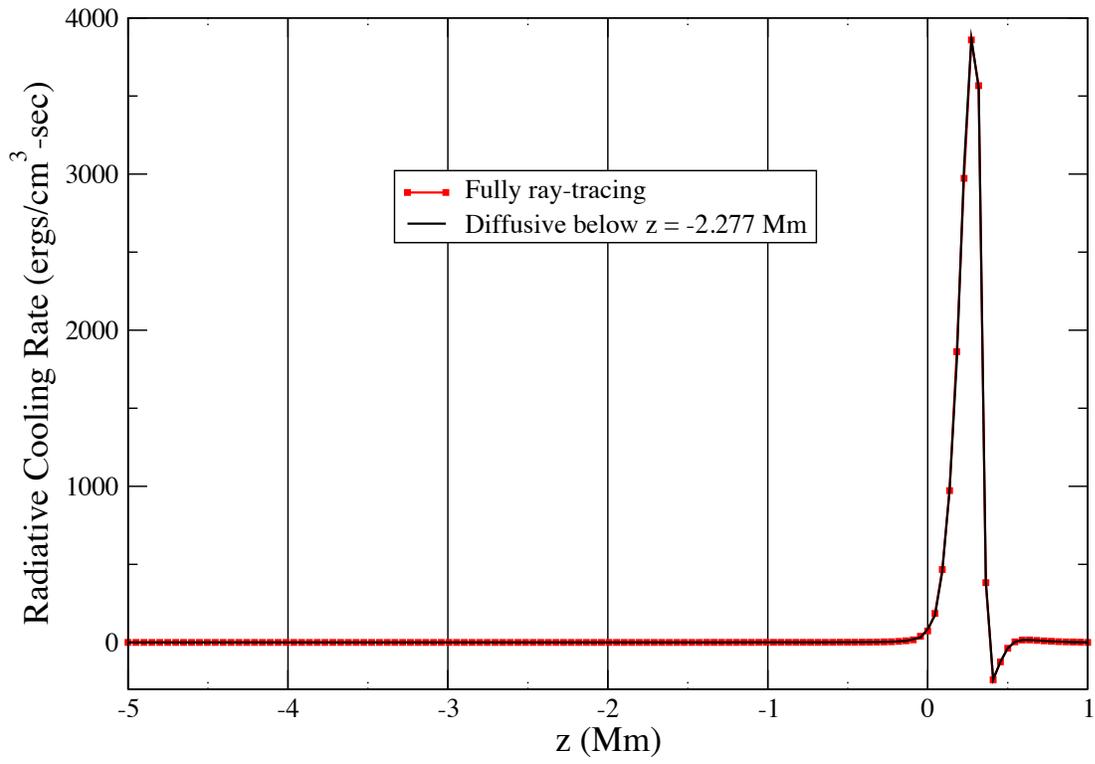}}
\caption{Comparison of radiative transfer calculations over the full domain; red line with squares: full ray-tracing; black line: diffusive radiation treatment in the optically thick region (below $z=-2.277$ Mm).}
\label{fig:radcomp_full}
\end{figure}

\begin{figure}[h]
\centerline{\includegraphics[width=\linewidth]{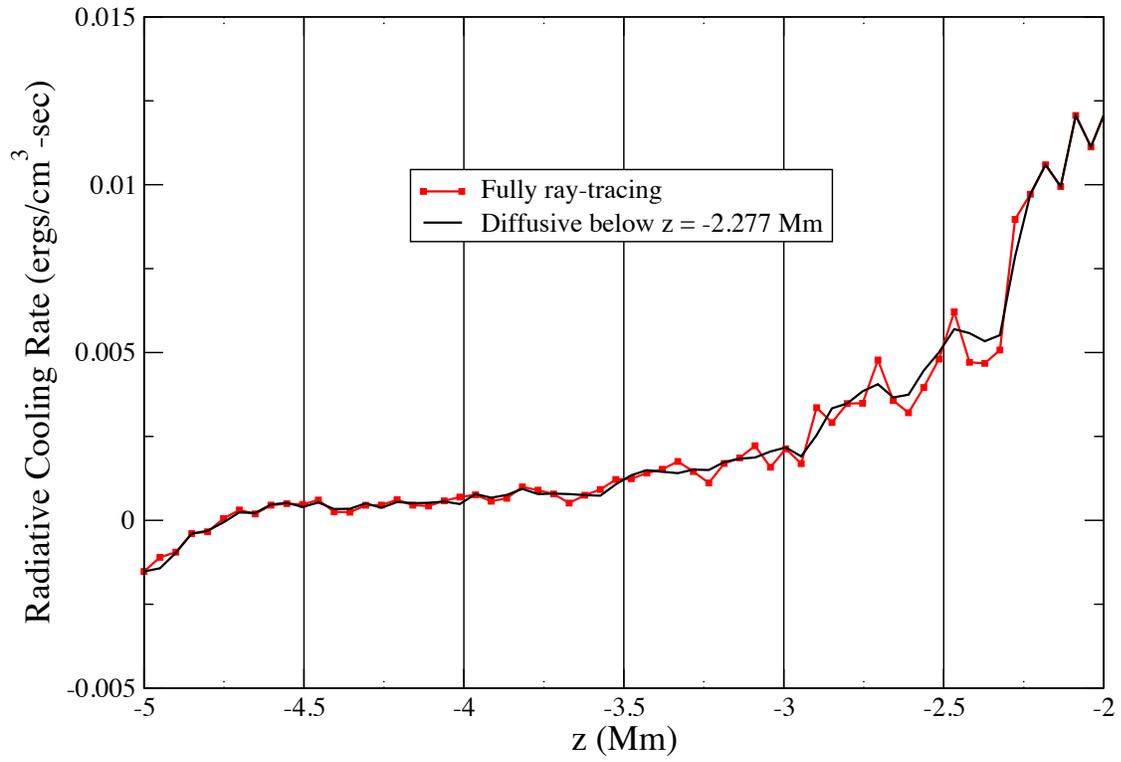}}
\caption{Comparison of radiative transfer calculations below $z=-2$ Mm: same curves as in Fig. \ref{fig:radcomp_full}.}
\label{fig:radcomp_zoom}
\end{figure}

Figures \ref{fig:radcomp_full} and \ref{fig:radcomp_zoom} compare the local radiative cooling obtained for these two simulations. The profiles are at a single time and are averaged over $x$-$y$ planes.  They clearly agree very well over the full domain in $z$ (Fig. \ref{fig:radcomp_full}), and, in the optically thick region below $z=-2$ Mm (Fig. \ref{fig:radcomp_zoom}), agreement is also good, though it is worth noting that the profile is smoother for the diffusive method, most likely because it does not involve a discrete angular quadrature; in any case, the values are very close between the two approaches.

\section{Simulations of Stellar Magnetoconvection}

In this Section we present some results of our simulations of solar and stellar magnetoconvection. While the main goal is to demonstrate the code's capabilities for understanding the dynamics of near-surface turbulent convection, the  results give important insights into stellar surface dynamics and magnetic effects.  
 
\subsection{Effects of Numerical Grid Resolution}

\begin{figure}[h]
\centerline{\includegraphics[width=\textwidth]{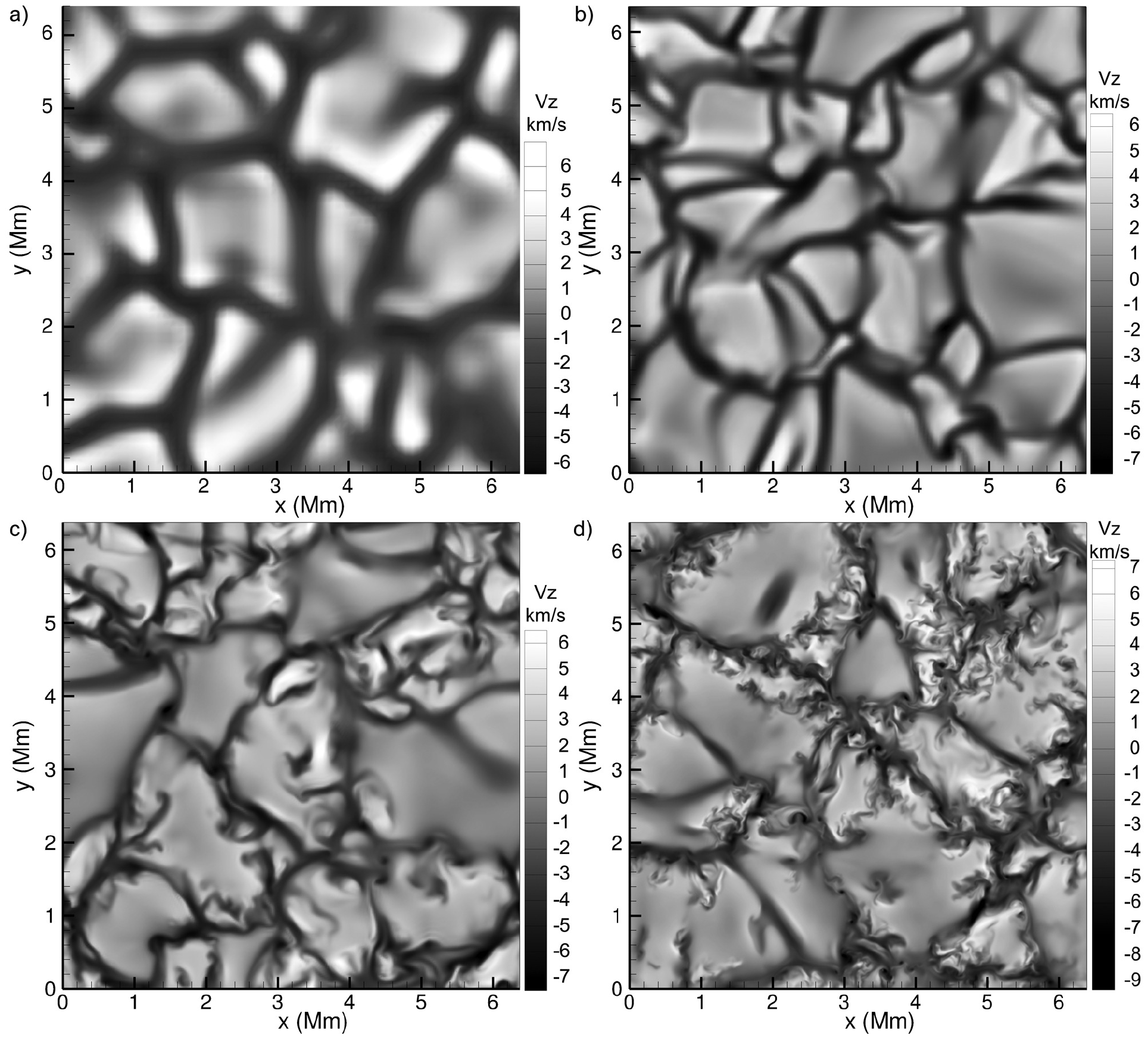}}
\caption{Snapshots of the simulated solar granulation (vertical velocity shown) at the photosphere for different resolutions:  $a$) 100~km, $b$) 50~km, $c$) 25~km and $d$) 12.5~km. }
\label{fig:ik1}
\end{figure}

\begin{figure}[h]
\centerline{\includegraphics[width=0.6\textwidth]{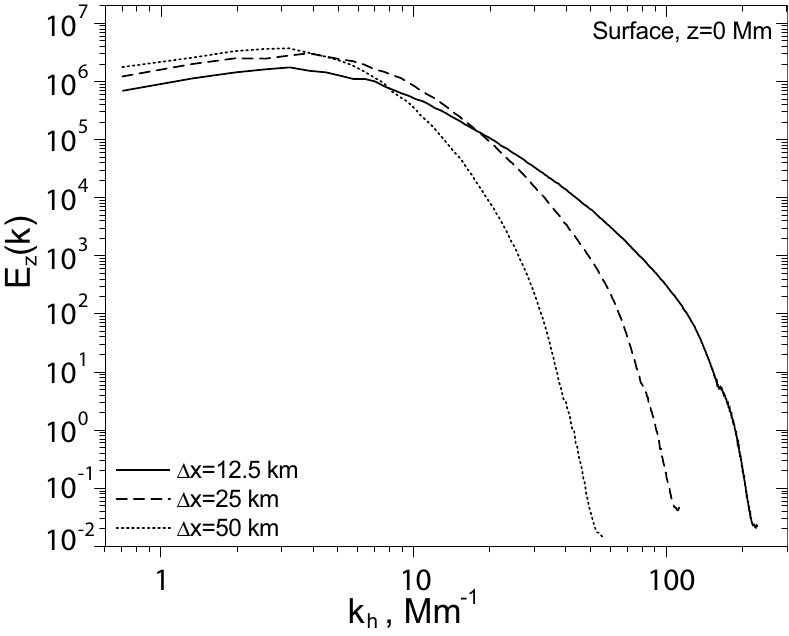}}
\caption{Turbulent energy spectra of the vertical velocity of solar convection at the photosphere for $\Delta x =$ 50, 25 and 12.5~km.
}
\label{fig:ik2}
\end{figure}

Despite substantial growth in computing power, our computational capabilities for 3D simulations are still quite limited, and it is important to investigate the effects of numerical grid spacing on the resolution of the important turbulent convective structures, such as granulation. Figure \ref{fig:ik1} presents snapshots of the vertical velocity at the solar surface for simulations with grid spacings of 100, 50, 25, and 12.5 km. The results show the granulation structure, which has a characteristic size of about 1~Mm and consists of relatively slow upflows occupying most of the area and fast downdrafts concentrated in the intergranular lanes. The primary effect of the decreasing grid spacing is in resolving the small-scale  structures and dynamics in the intergranular lanes; the primary granulation structure is well-resolved at all resolutions. This is also evident from  the turbulent energy spectra shown in Figure ~\ref{fig:ik2}. The energy spectra show that resolving small-scale turbulence may significantly reduce the large-scale energy of the granulation. This effect has to be taken into account in stellar simulations which are generally performed with relatively low resolution.

\begin{figure}[h]
\centerline{\includegraphics[width=0.6\textwidth]{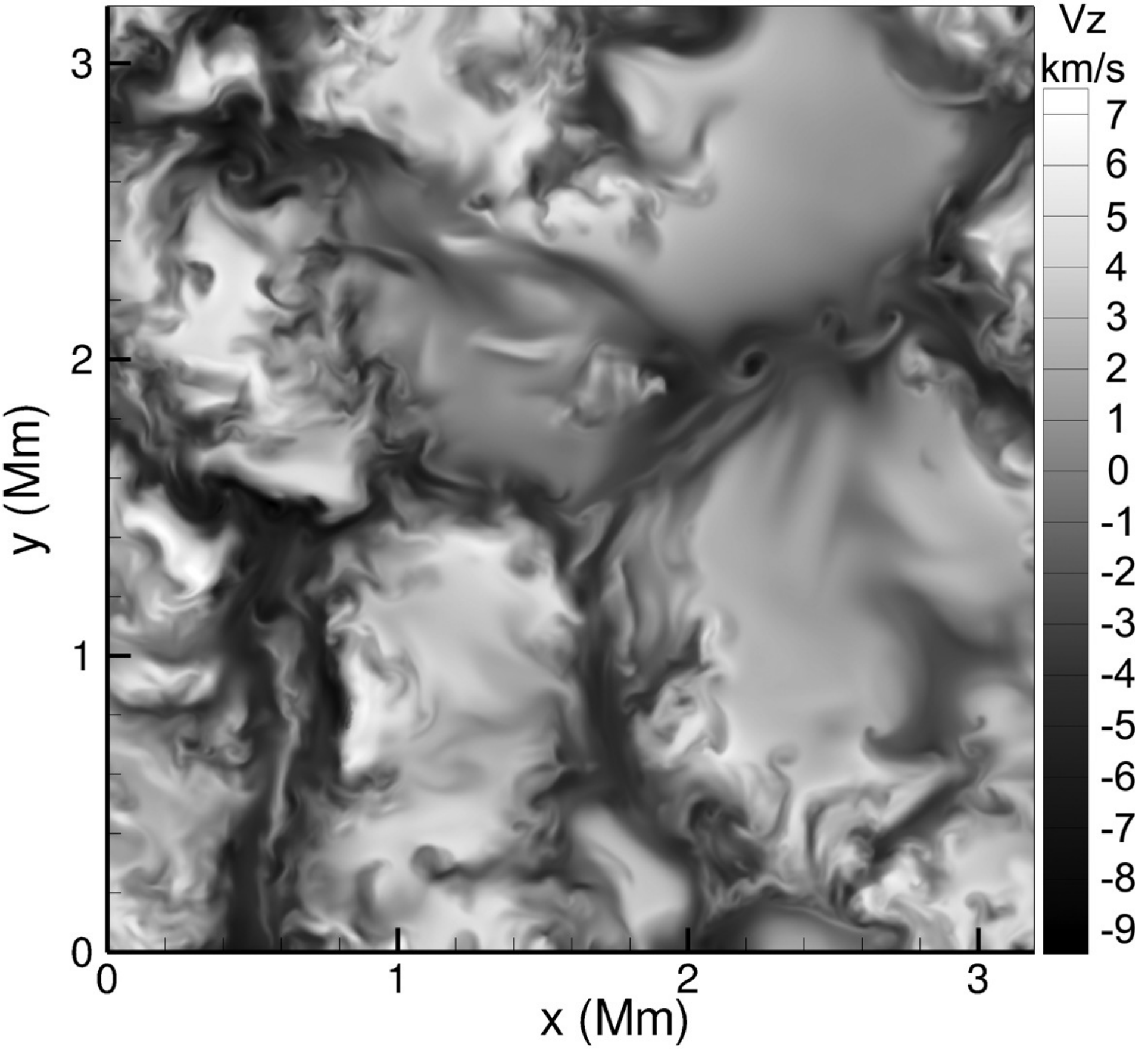}}
\caption{High-resolution simulations ($\Delta x = 6.25$~km) of solar convection revealing a very inhomogeneous distribution of the vertical velocity at the photosphere.
}
\label{fig:ik3}
\end{figure}

The small-scale dynamics of the intergranular lanes of solar convection, illustrated in Figure~\ref{fig:ik3}, is associated with shearing flows and plays a critical role in the formation of tornado-like vortex tubes \citep{Kitiashvili2012}, excitation of acoustic waves \citep{Kitiashvili2011}, creation of local dynamos \citep{Kitiashvili2013a}, etc. For example, powerful vortex tubes are evident in Fig.~\ref{fig:ik3} as rounded structures, $\sim 100$~km across, with dark cores corresponding to nearly supersonic downdrafts.  Undoubtedly such shearing and twisting flows in the intergranular lanes are also extremely important on other stars.  These effects will be a topic in future investigations using the StellarBox code.

\subsection{Structure of Granulation on the Sun and Moderate Mass Stars} 

\begin{figure}[h]
\centerline{\includegraphics[width=0.9\textwidth]{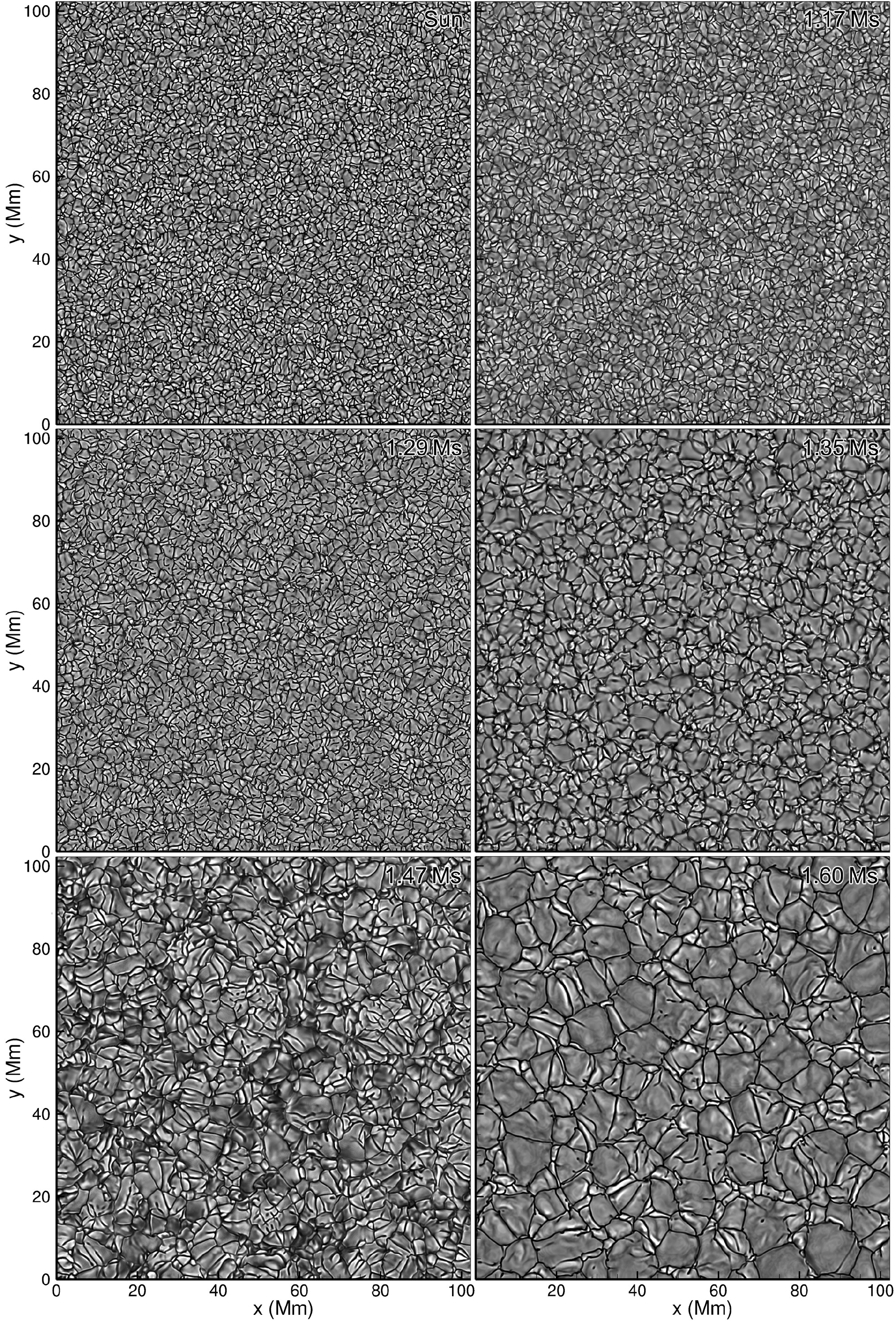}}
\caption{Variation of the scales of granulation for main-sequence stars with increasing stellar mass; from top, left to right: 1~M$_\odot$, 1.17~M$_\odot$, 1.29~M$_\odot$, 1.35~M$_\odot$, 1.47~M$_\odot$, 1.60~M$_\odot$. Distribution of the vertical velocity is plotted for a range of $\pm 6$~km/s in all plots. }
\label{fig:ik4}
\end{figure}

As stellar mass increases, the outer convection zone shrinks, and turbulent motions become more vigorous because of the increased energy flux. The granulation structure also changes quite significantly. Figure~\ref{fig:ik4}  shows the distribution of the vertical velocity on the surface of the Sun and five main-sequence stars with masses of   1.17~M$_\odot$, 1.29~M$_\odot$,  1.35~M$_\odot$,  1.47~M$_\odot$, and 1.60~M$_\odot$. The simulations were performed for a computational domain of $100\times 100$~Mm horizontally, $40-50$~Mm in depth, and $1$~Mm above the photosphere. The initial conditions in each case are for standard zero-age main-sequence models calculated for the solar composition using the CESAM code \citep{Morel1997}.
The initial conditions for the solar simulations are constructed by combining
the standard model S (\citep{Christensen-Dalsgaard1996}) and the VAL model
of the solar atmosphere (\citep{Vernazza1973}).
These hydrostatic initial conditions are perturbed by random velocity fluctuations to initiate convective motion, and simulation runs are continued for several hours of stellar time until statistically stationary conditions are achieved. No magnetic fields were included in these simulations.

The simulation results show that the characteristic size of granulation increases from $\sim$~1~Mm in the case of the Sun to more than 10~Mm for the  1.60~M$_\odot$ A-type star. For this latter type of star, standard stellar evolution theory using a mixing-length model predicts that the outer convection develops only in the hydrogen and helium ionization zones, separated by a stable layer. However, the simulations show that these layers are mixed, creating a shallow convection zone with a well-defined granulation pattern.  For reference, the pressure scale heights for the Sun and the 1.17~M$_\odot$, 1.29~M$_\odot$,  1.35~M$_\odot$,  1.47~M$_\odot$, and 1.60~M$_\odot$ stars are approximately 140, 173, 236, 267, 270, and 359~km, respectively.

An interesting question concerns the large-scale organization of granulation, observed on the Sun as meso- and super-granulation. Our results show that the solar granulation tends to form rather irregular clusters of several granules. However, the simulations do not explicitly show any large-scale (20-30~Mm) patterns which could correspond to supergranulation, at least for zero rotation and the simulation domain dimensions tried so far (the domain extended to 20~Mm deep for stars with masses $\leq$ 1.35~M$_\odot$ and to approximately 50~Mm deep for heavier stars). This means that some essential physics may be missing in these simulations, e.g., large-scale magnetic fields or rotation. This question requires further investigation. Mesoscale clustering can be seen in the more massive stars. It is particularly pronounced in the 1.47~M$_\odot$ case, in which large-scale convection cells become crossed, ``shredded'', by a series of aligned intergranular lanes. This type of granular instability also exists in solar granulation \citep{Kitiashvili2012} but is substantially less pronounced. The shredding process is accompanied by generation of intense acoustic waves which are seen as diffuse darker patches in the snapshots.

These results demonstrate the capabilities of the StellarBox code in simulating turbulent stellar convection and reveal very complex multi-scale convective structures.  A more detailed understanding of their dynamics will be a goal of our future studies. 

\subsection{Magnetic Field Structuring}

\begin{figure}[h]
\centerline{\includegraphics[width=\textwidth]{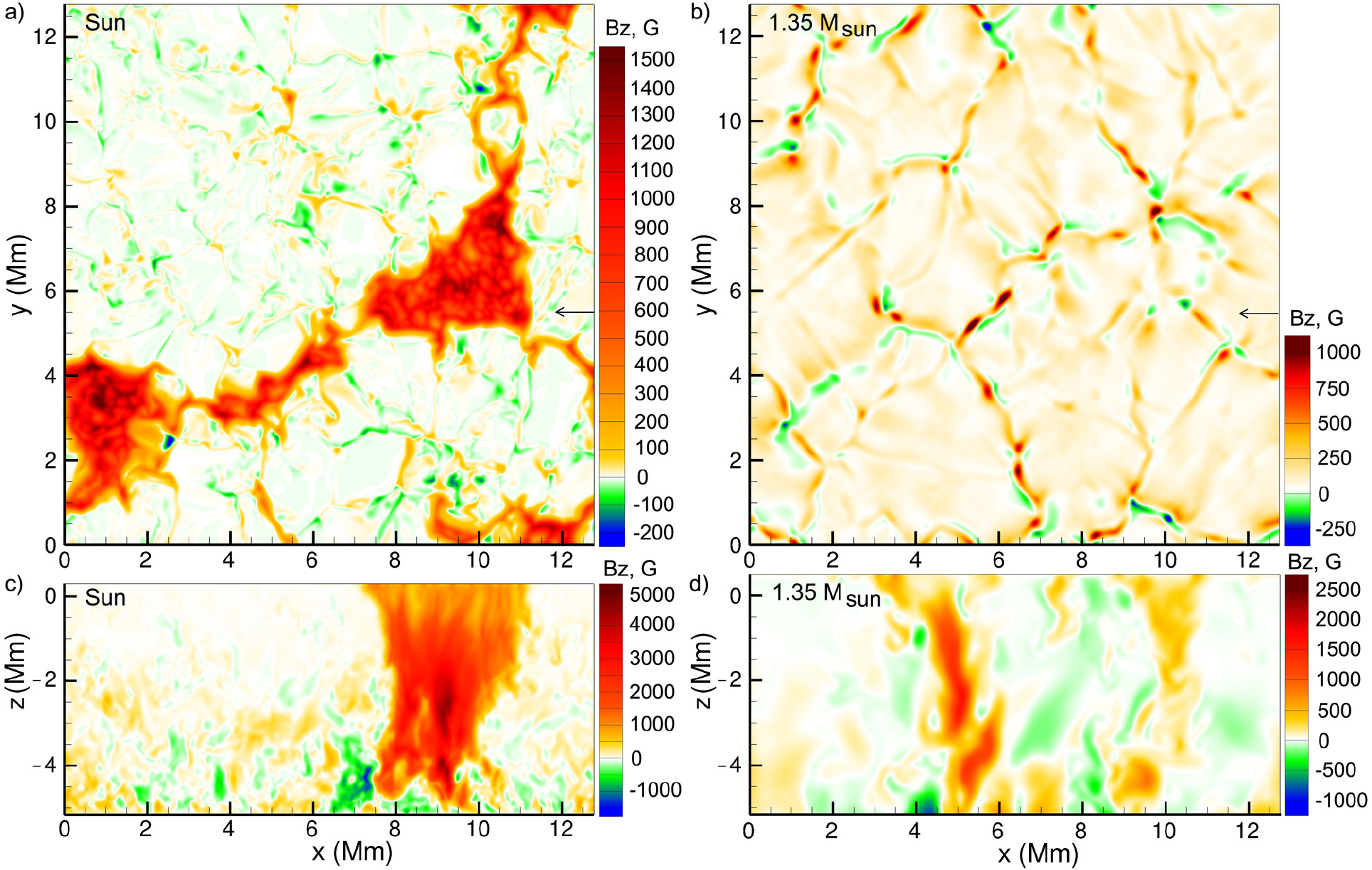}}
\caption{Distribution of magnetic field in the photosphere from simulations with an initially uniform vertical 100~G magnetic field: a) in the Sun, and b) in a 1.35M$_\odot$ star. Panels c) and d) show corresponding vertical cuts along the $x$-axis at the locations indicated by the arrows in a) and b). The fields shown are typical of the statistically stationary state.}
\label{fig:ik5}
\end{figure}

To illustrate the effects of magnetic fields on near-surface stellar convection, we present simulation results following the injection of a uniform 100~G vertical magnetic field into a fully developed convection layer. The boundary conditions conserve the mean vector magnetic field but do not prescribe any field structure. The domain is $12.8\times 12.8$~Mm horizontally, 5.5~Mm in depth below the photosphere, and 0.5~Mm above the photosphere. Results for the Sun and a 1.35~$M_\odot$ star are shown in Figure~\ref{fig:ik5}. In the case of the Sun, the initially uniform magnetic field becomes concentrated into compact, self-organized, 3-4~Mm wide pore-like magnetic structures maintained by strong dowdrafts. This process of spontaneous formation has been previously described in detail by \cite{Kitiashvili2010} for simulations in a smaller, $6\times 6$~Mm domain. The new simulations confirm these results and show that the larger domain did not lead to formation of a larger structure.  Instead, two compact structures of a similar size were formed, indicating that the structures are independent of the simulation domain size. What determines the scale of these self-organized magnetic domains has not been established. Curiously, the simulations for the 1.35~$M_\odot$ star did not result in formation of such large-scale structures, but instead the magnetic field became concentrated in small-scale patches in the intergranular lanes. In addition, the magnetic field formed diffuse patches of 100-200~G in the bodies of the stellar granules, whereas the magnetic fields inside the solar granules were much weaker.   The reason for this difference is also unclear. Nevertheless, this indicates that the background (`basal') magnetic field may have quite different structures on different types of stars. From the numerical point of view, our simulations have shown that, for simulating the process of spontaneous structure formation, the grid spacing should be 25~km or smaller. For coarser grids the structures do not form.  This indicates the importance of accurate simulation of the flow dynamics in intergranular lanes.

\section{Summary}
We have presented the basic features and some test results of the 3D radiative MHD code `StellarBox'. The code is designed to accurately simulate turbulent magnetoconvection processes in solar and stellar envelopes. It is based on a high-order Pad\'e finite-difference scheme and implements subgrid-scale Large-Eddy Simulation (LES) turbulence modeling. This is expected to provide a more accurate description, as compared to using more ad~hoc turbulence models or lower-accuracy numerics, of complex physical phenomena in the highly turbulent radiating plasma in the Sun and other stars.  Such phenomena include the formation and dynamics of surface granulation, large-scale convective structures, excitation of acoustic, gravity, and MHD waves, magnetic self-organization, etc. The code has been carefully tested using standard CFD and MHD solutions and shows good accuracy and robustness. 

The code's capabilities were demonstrated by performing simulations of the upper  convection zones of the Sun and several moderate mass stars, first without magnetic field and then with an imposed, initially uniform, vertical  magnetic field. The solar convection simulations were performed for different numerical grid spacings, from 6.25~km to 100~km. The results show that, while the coarse, 100~km grid is sufficient for resolving the granulation structure, the rich dynamics of  intergranular lanes can only be resolved with smaller grid spacings, e.g. 12~km. The importance of such resolution is revealed by the observation that the intergranular lanes are a source of strong, almost supersonic shearing flows and compact ($\sim 100$~km across) vortex tubes, which play important roles in the formation of magnetic structures. 

For an initial comparative analysis of stellar convection, we performed large-scale ($100\times 100\times 40$~Mm) simulations for six main-sequence stars with masses from 1.0~M$_\odot$ to 1.60~M$_\odot$. The results showed a substantial increase in the granulation size with stellar mass, from 1 to 20 Mm. Further, the granulation for stars more massive than the sun was often clustered on mesogranulation scales (on the scale of several granules). 

It was found that magnetic field effects can be quite different in stars of different classes. The simulations showed that, in the solar mass case, an initially uniform magnetic field forms self-organized, stable `pore-like' structures  of the size of several granules. But, in the case of more massive stars, such structures are not formed, and the magnetic field is distributed in the intergranular lanes in the form of small kilogauss  magnetic flux tubes.

In conclusion, 3D radiative MHD simulations, which are becoming more and more feasible on modern supercomputer systems, allow us to simulate stellar magnetoconvection with a great degree of realism and provide an important tool for understanding the complex physics of turbulent stellar envelopes. 

\section*{Acknowledgments}
The work was partially supported by NASA grants NNX09AJ85G
and NNX14AB70G.

\bibliographystyle{apj}

\newpage

\end{document}